\documentstyle[eqsecnum,aps]{revtex}
\draft
\newfont{\msam}{msam10}
\def\zid{\hbox{{1}\kern-.25em\hbox{l}}}
\begin{document}
\title{Mod.Phys.Lett.A, in press\quad hep-th/9305067 \quad UW/PT-93-05
\quad ANL-HEP-CP-93-39 \\
.\phantom{forcing title to leave a mostly blank blank blank blank blank
blank blank blank blank blank blank line}\\
.\phantom{forcing title to leave a mostly blank blank blank blank blank
blank blank blank blank blank blank line}\\
Thermodynamic q-Distributions That Aren't }
\author{Stamatis Vokos}
\address{Department of Physics, FM-15\\
University of Washington,\\
Seattle, WA 98195}
\author{Cosmas Zachos \thanks{Electronic mail: zachos@hep.anl.gov} }
\address{High Energy Physics Division,\\
  Argonne National Laboratory,\\
  Argonne, IL 60439-4815, USA}
\date{May 1993}

\maketitle

\begin{abstract}
Bosonic q-oscillators commute with themselves and so their free distribution
is Planckian. In a cavity, their emission and absorption rates may grow or
shrink---and even diverge---but they nevertheless balance to
yield the Planck distribution via Einstein's equilibrium method,
(a careless application of which might
produce spurious q-dependent distribution functions).
This drives home the point that the black-body energy distribution is not a
handle for distinguishing q-excitations from plain oscillators.
A maximum cavity size is suggested by the inverse critical frequency of
such emission/absorption rates at a given temperature, or a maximum temperature
at a given frequency.
To remedy fragmentation of opinion on the subject, we provide some discussion,
context, and references.
\end{abstract}
\pacs{11.30.Ly, 12.90.$+$b, 05.30.Jp, 03.65.Fd}

\widetext                                
\section{ Introduction}
The excitations of various systems, such as nonlinear masers,
interacting-magnon-cavities, etc.\cite{flortom,cstg,bubo,chailatest} are
often modelled through q-oscillators, most usefully for cyclotomic values
of the deformation parameter. In some contrast, for generic values of the
deformation parameter, as always in quantum algebras, q-oscillators amount to a
mere change of coordinates\cite{cz1,cgz}, advantageous or not, for describing a
given physical situation.

This is a summary of our unpublished April 1992 notes.  Subsequently, we
appreciated that the paradox discussed below had also already been resolved
correctly by \cite{bubo,agar} and, incorrectly,  by \cite{leeyu} and a number
of other authors. Because the bibliography is so fragmented, however, and
because the same points are repeatedly being discussed, often clarified, but
often confused despite the extant bibliography, we decided to summarize some
conventional thinking below, and to provide some references. For another broad
survey of the subject, the reader may consult \cite{chais}.

Bosonic q-oscillators are bosons (they commute with each other), and thus
free collections of them are described by the conventional Bose-Einstein
distribution. On the other hand, it is possible to see (e.g.\cite{polya,yan})
that for special values of the deformation parameter $q$ the deformed bosons
{\em discontinuously} start to exclude each other and collapse to fermions.
This has
suggested the notion that the {\em interactions} which are naturally
systematized by the q-oscillator formalism, and which thus suggest q-dependent
partition functions and distributions, should somehow interpolate between the
Fermi and the Bose-Einstein distribution.

Unfortunately, an easy, but specious, way to readily produce q-dependent
density distribution functions is via \cite{einstein}'s celebrated method of
balancing the equilibrium absorption and emission rates from a black-body wall
of a cavity without any direct reference to the hamiltonian of the gas of
q-modes in the cavity. A {\em naive} application of this technique would appear
to yield q-dependent distributions for free modes, which, however, as
mentioned, are mere bosons. Nevertheless, this paradox is resolved by correct
taking of thermal averages, and the conventional Planck-Bose-Einstein
distribution ensues, without much bearing on the coordinates used, as expected.

The minimum value of the frequency for which the absorption/emission rate of
quasi-modes  involved converges suggests an inverse maximal cavity size
$1/\omega_c$, below which q-oscillators are not a good description of the
system in question, at a given temperature; alternately, a maximum temperature
for a given frequency; but, depending on the physics problem addressed,  it may
also indicate a singularity of the coordinate description employed.

\section{ brief review of q-Heisenberg algebras}

The $q$-Heisenberg algebra, which is ultimately traceable to unpublished work
of
Heisenberg through \cite{heis}, is derivable though a map from SU(2)$_q$.

 Consider the following formal contraction of SU(2)$_q$
\begin{equation}    [J_0,J_+]=J_+  ~~~~~~~~~~~~~ ~~ [J_+,J_-]={1\over 2}~
(q^{2J_0} -q^{-2J_0})/ (q-q^{-1})~~~~~ ~~~~~~~~~ [J_-,J_0]=J_-~,
 \end{equation}
namely\cite{ce} and \cite{ng} (contrast to \cite{celeg}):
\begin{equation}
b\equiv q^{J_0} J_- \sqrt{2(q-1/q)}~,~~~~~~~~~~~~~~~~~
b^{\dagger}\equiv J_+ q^{J_0} \sqrt{2(q-1/q)} ~,
\end {equation}
so that
\begin{equation}
[J_0,b^{\dagger}]=b^{\dagger}    ~,~~~~~~~~~~~~~~~~~~~~~~~~~ [b,J_0]=b ~,
\end {equation}
and hence
\begin{equation}
b b^{\dagger} -q^2  b^{\dagger} b = \zid -q^{4J_0}.
\end {equation}
The last term on the r.h.s.\ vanishes e.g.\ for $|q|>1, ~~J_0<<0$ in an
infinite-dimensional representation (Schwinger's contraction), to yield
the $q$-oscillator algebra
\cite{cigler,kurysh,jannussis,macfarlane,bied,kuldam}
of

\noindent $\bullet$ {\sf TYPE} $b$:
\begin{equation}
b b^{\dagger} -q^2  b^{\dagger} b =\zid ~.
\end {equation}

The conventional (Bargmann holomorphic) realization
\cite{cigler,alvgaum,ruegg,flortom,bracken} for this algebra is
$b^{\dagger}=x$ and
$b=D_{q^2}$, where $D_{q}$ is the quantum derivative, i.e.~the slope of the
chord to the graph of a function between $x$ and $qx$:
\begin{equation}
D_{q}f(x)\equiv { f(qx)-f(x)\over x(q-1)}~.
\end {equation}
(For the spacetime, q-Hermite polynomial, realization of the
q-oscillator system see \cite{atak,flor,jvdj,min}).\footnote{
For q-coherent and q-squeezed states, see
\cite{arik,jannussis,bied,bracken,nelson,celsqueez}.}

The number operator
\begin{equation}
N\equiv \ln{\Bigl(1+(q^2-1)b^{\dagger} b\Bigr)}/\ln{q^2}~~~~~~~~~~~~
\hbox{so~that}~~~~~~~~~~~~~~~~[N]_b\equiv {q^{2N}-1\over q^2 -1}= b^{\dagger} b
\end {equation}
may be introduced \cite{macfarlane}, hermitean for real $q$, s.t.\
\begin{equation}
[N,b^{\dagger}]=b^{\dagger}  ~,~~~~~~
[N,b]=-b~,~   ~~~ [b,b^{\dagger}] =[N+1]_b - [N]_b=q^{2N}~.
\end {equation}
This $q$-oscillator algebra can then be mapped to the alternate form of

\noindent $\bullet$ {\sf TYPE} $\alpha$:
\begin{equation}
\alpha=q^{-N}b,~~~    ~~~    \alpha^{\dagger} =b^{\dagger} q^{-N}~~~ ~~~ ~~~
\Longrightarrow ~~~    ~~~
{}~~~[\alpha ,\alpha^{\dagger}]= q^{-2N}  ~,
\end {equation}
with
\begin{equation}
\alpha^{\dagger} \alpha = (1-q^{-2N} )/(1-q^{-2})\equiv [N]_{\alpha}~,
{}~~~~~~~~~\alpha \alpha^{\dagger} =[N+1]_{\alpha}~,
{}~~~~~~~~~\alpha \alpha^{\dagger} -q^{-2} \alpha^{\dagger} \alpha = \zid.
\end {equation}
Note the reflection to  the previous q-oscillator type via $q\mapsto 1/q$.
Alternatively, map to the popular form of

\noindent $\bullet$ {\sf TYPE} $a$:
\begin{equation}
a=q^{-N/2}b,~~~~~~        a^{\dagger} =b^{\dagger} q^{-N/2}~~~    ~~~
\Longrightarrow  ~~~    ~~~aa^{\dagger}-qa    ^{\dagger} a = q^{-N}~,
\end {equation}
with
\begin{equation}
a^{\dagger} a =[N]_a\equiv {q^{N}-q^{-N}\over q -1/q}
{}~~~ ~~~ \Longrightarrow  ~~~ [a,a^{\dagger}] =[N+1]_a - [N]_a .
\end {equation}
An alternate, less symmetric (non-hermitean) map
\begin{equation}
a=q^{-N}b,~~~~~~        ~~~~~~        a^{\dagger} =b^{\dagger} ~,
\end {equation}
which leads to the same algebra, produces the holomorphic realization
 $a^{\dagger}=x$ and
$a={\cal D}_{q}$, where now
\begin{equation}
{\cal D}_{q}f(x)\equiv { f(qx)-f(x/q)\over x(q-1/q)}~.
\end {equation}

These three types are then largely equivalent, and likewise equivalent to
classical oscillators for generic $q$.
Deforming functionals for the $q$-Heisenberg algebra of type $\alpha$ are
\cite{cigler,kurysh,jannussis,polya,song}:
\begin{equation}
a ^{\dagger} = \sqrt{{[N]_a\over N}}  A^{\dagger} ,~~~
{}~~~~~~ ~~~~~~ ~~~~~~ a=A\sqrt{{[N]_a\over N}} ~,
\end {equation}
where the classical oscillator algebra is
\begin{equation}
[A,A^{\dagger}]=\zid  ~,~~~~~~ ~~~~~~ ~~~~~~~~~~N=A ^{\dagger} A~,
\end {equation}
consistent with the above.

Mutatis mutandis,
\begin{equation}
\alpha ^{\dagger} = \sqrt{{[N]_{\alpha}\over N}}  A^{\dagger} ,~~~
{}~~~~~~ ~~~~~~ ~~~~~~ \alpha=A\sqrt{{[N]_{\alpha}\over N}} ~,
\end {equation}

\begin{equation}
b^{\dagger} = \sqrt{{[N]_b\over N}}  A^{\dagger} ,~~~
{}~~~~~~ ~~~~~~ ~~~~~~ b=A\sqrt{{[N]_b\over N}} ~.
\end {equation}

It then follows directly that these q-algebras dictate, for $N | n  \rangle=
n| n  \rangle$:
\begin{equation}
\langle n+1  | a^{\dagger}   | n  \rangle= \sqrt{[n+1]_a}
\end {equation}
and its hermitean conjugate:
\begin{equation}
\langle n-1  | a| n  \rangle= \sqrt{[n]_a}~,
\end {equation}
and similarly
\begin{equation}
\langle n-1  | b| n  \rangle= \sqrt{[n]_b}~,  \qquad  \qquad
\langle n-1  | \alpha | n  \rangle= \sqrt{[n]_{\alpha}}~.\end {equation}

It follows by inspection of these deforming maps that all these
excitations, $a^{\dagger},A^{\dagger},b^{\dagger},\alpha^{\dagger}$,
commute among themselves, and so do their annihilation operators,
like good bosons. The deformation maps detailed above are invertible for
generic $q$,
i.e.~not a root of unity, and thus they may be effectively regarded as partial
resummation of a perturbation series\footnote{However, for $q$ a root of
unity, powers of the creation or annihilation operators will
vanish\cite{polya,yan}, and for $q=i$ fermions result, as the square of two
creators or annihilators vanishes: observe that $[2]=0$ for all three types.
For a generic discussion in purely algebraic terms, see \cite{zachos}.}.

\section{Naive quasi-mode distributions}

The hamiltonians formed by the above q-boson mode oscillators are as
infinitely diverse as the  exotic functions of the number operators $N$ (with
integral eigenvalues)---or the [$N$]'s---one could think of. They are {\em
interacting}
with the exception of the plain $N$, the provider of the truly free,
 linearly-spaced, spectrum\footnote{ Since any and all oscillators are
deformable to each other for {\em generic q}, the {\em type} of oscillators
used says little about the spectrum: via
the above  deforming functionals, it is evident that any hamiltonian may be
represented in terms of any type of oscillators, and convenience is the
dominant consideration. Nevertheless, some authors
appear to suggest that the aesthetic simplicity of a hamiltonian bilinear in
some type of oscillator may go beyond a trivial coordinate-system statement
to somehow render it ``chosen", and its (interacting) spectrum somehow
``basic".}. Via Bose statistics, these determine
partition functions, and whence q-dependent spectral density distributions.
E.g. \cite{martin,bubo,nesko},  \cite{jannussis,ce,flortom,manko,su,shanta}.

Nevertheless, for {\em free} mode-gas hamiltonians, given the absorption and
emission rates from a black-body wall of a cavity, \cite{einstein} indicates
how to compute density distribution functions directly from the condition of
equilibrium. Specifically, let absorption and emission (spontaneous and
induced) rates for q-phonons of energy $\hbar \omega$ be $C_a$, $C_e$. The
thermalized cavity wall molecules populations are controlled by the Boltzmann
distribution $\exp(-E/kT)$, so that the relative population ratio of
emitting to absorbing molecules is $\exp(-\hbar \omega /kT)$. Incorporating,
in the spirit of Wien's law,  $\hbar /kT$ into $\omega$, for an equilibrium
distribution of excitations in the cavity,
\begin{equation}
C_a= C_e ~ e^{-\omega}
\end {equation}
holds.
For ordinary photons, the QED interaction hamiltonian directly dictates
$C_e\propto  \langle n\rangle+1$   and    $C_a\propto \langle n\rangle$,
leading to the Bose-Einstein distribution of the  Planck law:
\begin{equation}
\langle n(\omega)\rangle= {1\over e^{\omega} -1}  ~.
\end {equation}
For the above quasi-modes, however, eqs.(2.19-21), and a linear hermitean
interaction for such deformed q-boson excitations lead to
\begin{equation}
C_e\propto \langle [n+1] \rangle, ~~~~~~~~~~~~~~~~~~~~~~C_a\propto
\langle [n]\rangle.
\end {equation}

\noindent {\bf An opportunity to err.} For the conventional boson case, $n$
and $\langle n \rangle$ enjoy exactly the same algebraic status, by linearity.
Not so for the q-case, though: if thermal averages were overlooked,
 i.e. if the averages of these hyperbolic functions were the same
functions of the averages $\langle n\rangle$, then one could ignore the
difference  between  $n$ and $\langle n\rangle$, and proceed as before.
The equilibrium reversibility condition would then yield the spurious
distributions:

\begin{equation}
n_a(\omega)= {1\over \ln (q^2)} \ln{1/q -e^{\omega} \over q-e^{\omega}}~,
\qquad \qquad
n_b(\omega)= {1\over \ln (q^2)} \ln{1 -e^{\omega} \over q^2-e^{\omega}}~,
\qquad \qquad
n_{\alpha}(\omega)= {1\over \ln (q^2)} \ln{q^{-2} -e^{\omega}
\over 1-e^{\omega}}~.
\end {equation}

Naturally, all three distributions reduce to the classical one $n(\omega)$ as
$q\rightarrow1$. Note that as $q\rightarrow 1/q$, $~n_a$ remains invariant,
whence $n_a$ is real for $q=e^{i \phi}$, a pure phase; while $n_b
\leftrightarrow n_{\alpha}$.\footnote{ For real $q$, not all infrared
singularities of the above distributions are at $\omega=0$. In particular, for
$q>1$,

$n_{\alpha}$ only has a singularity at $\omega_c=0$, but

$n_b$ diverges at $\omega_c=2\ln q$, and goes imaginary for lower frequencies;

$n_a$ diverges at $\omega_c=\ln q$, and goes imaginary for lower frequencies.

\noindent For $q<1$,

$n_b$ only has a singularity at $\omega_c=0$, but

$n_{\alpha}$ diverges at $\omega_c=-2\ln q$, and goes imaginary for lower
frequencies;

$n_a$ diverges at $\omega_c=-\ln q$, and goes imaginary for lower frequencies.}

For very large $\omega$,
\begin{equation}
n_a \sim e^{-\omega} {(q-1/q)\over \ln q^2},
\qquad \qquad
n_b \sim e^{-\omega} {(q^2-1)\over \ln q^2},
\qquad \qquad
n_{\alpha} \sim e^{-\omega} {(1-1/q^2)\over \ln q^2}~,
\end {equation}
amounting to a mere q-dependent frequency shift of the Wien-regime
distribution.

Multiplying by the excitation energy $\omega$ and the density of states in
the cavity of volume $V$ (in three dimensions)
yields the deformed spectral energy density
per-frequency-interval, generalizing the  law of Planck:
\begin{equation}
dU= {V (kT)^4\over \hbar^3 c^3 \pi^2}~~n_{a,b,\alpha} (\omega)~ \omega^3 ~
d\omega,
\end{equation}
where the rescaled definition of frequency must be borne in mind.

But this would be highly {\em paradoxical}, as the free quasi-modes commute
with each other,  and so should obey the Bose-Einstein distribution, which is
the distribution dictated directly by the partition function of a
non-degenerate equally-spaced  spectrum, and is independent of $q$.

\section {non-trivial thermal averages: A free system of commuting quasimodes
is Einstein-Planck distributed}

Of course, the functions thermally averaged were not linear, and hence not
trivial. \cite{bubo,agar} compute these averages correctly. In contrast to
the linear case, explicit knowledge of the (trivial) spectrum is now necessary.

One only needs to recognize that, for free
excitations  (whence a linear, additive, spectrum),
\begin{equation}
\langle [N]_b\rangle= {\sum_n  e^{-n\omega}  {q^{2n} -1   \over q^2 -1  }
    \over\sum_n  e^{-n\omega}  }   = 1/( e^{\omega}-q^2) =
{\langle N\rangle \over 1+(1 -q^2) \langle N\rangle} ~,
\end {equation}
where, for convenience,  we substitute for
$1/(e^{\omega} -1)=\langle N \rangle$ in the final expressions.

Moreover, in some generality, merely shift the argument of the exponential by
$\omega$, carry this one power outside, and compare to the above:
\begin{equation}
\langle [N+1]_b\rangle=e^{\omega} \langle [N]_b\rangle
= e^{\omega}/( e^{\omega}-q^2) =
{1+ \langle N\rangle \over 1+ (1 -q^2) \langle N\rangle}~ .
\end{equation}

As a consequence, even though the rates themselves are not Einstein, their
{\em ratio} is still the standard one. The black-body energy spectrum cannot
be used as a handle for distinguishing quasi-modes from plain oscillator
excitations. Emission and absorption accelerate wildly
to {\em infinity} at  the critical value of the frequency $\omega_c=2\ln ~q$.
This does not signal an intrinsic condensation phenomenon, as the number
densities are well-defined at that frequency. The interaction mechanism simply
populates/depopulates high occupation numbers too rapidly. Reinstating the
factors  absorbed in the units of frequency, $\omega_c=2kT\ln q ~/\hbar$.
This minimum value of the
frequency specifies an {\em inverse maximal cavity size} $1/\omega_c$,
below which q-oscillators are not a satisfactory description of the
system under study, at a given temperature; or else a maximal temperature for
a given frequency; this may also amount to a singularity of the coordinate
description employed, of use or not in simulating nonlinear systems.

For $\alpha$-modes, substitute $q\mapsto 1/q$,
with the corresponding singularity for $q<1$ and the same distribution;
and for $a$-excitations,
\begin{equation}
\langle [N]_a\rangle={e^{\omega}-1\over (e^{\omega}-q)(e^{\omega}-1/q)}~,
\end{equation}
with $\omega_c=\mp \ln q$ for $q${\msam 7}1, and likewise the same
distribution.

Misinterpretation of emission and absorption rates as number density
distributions, however, \cite{leeyu}, eventuates in incorrect energy spectral
distributions\footnote{In fact, these references appear to
miss the singularity in the absorption rate discussed above, even though their
figure might hint at it.}. \cite{tusz,late} follow in much the same spirit to
reach spurious distributions of the type exemplified in the previous section.

Of course, for special cyclotomic values of the parameter $q$, the statistics
of
the particles do change, {\em abruptly}, as the Fock space is truncated
\cite{yan,polya}. But since the distribution functions do not depend on $q$,
they do not reflect these changes, which must be put in
by hand, taking account of state multiplicities.

In sharp contrast to the above discussion, for {\em interacting} hamiltonians,
the partition function genuinely depends on $q$, and hence
such q-distributions {\em are}. Nevertheless, they need not be associated to
a given deformation/coordinate description, except through considerations of
calculational convenience. The hyperbolic-sine-spectrum partition function
has not been summed explicitly, beyond graphing/expanding to
a few leading orders in $q-1$, \cite{martin,bubo,nesko,manko}; however,
this  is liable to miss interesting
nonanalytic structure, and thus possible condensation features that such
structure may underlie.

\acknowledgments
We wish to thank G. Bodwin, A. Abrikosov, Jr.,  and P. Freund for helpful
discussions. Work supported by the U.S.Department of Energy,
Division of High Energy Physics, Contract W-31-109-ENG-38, and the
DOE grant DE-FG06-91ER40614.

\end{document}